\documentclass[particles,article,accept,pdftex,moreauthors]{Definitions/mdpi} 

\firstpage{489} 
\pubvolume{7}
\issuenum{501}
\articlenumber{0}
\pubyear{2024}
\copyrightyear{2024}
\externaleditor{Academic Editor: Ilídio Lopes, Violetta Sagun and Laura Sagunski}
\datereceived{24 May 2024} 
\daterevised{7 June 2024} 
\dateaccepted{11 June 2024} 
\datepublished{13 June 2024 } 
\hreflink{https://doi.org/10.3390/particles7020028} 

 \usepackage{soulutf8}
\usepackage{amsfonts}
\usepackage{microtype}
\usepackage{siunitx}
\usepackage{subcaption}
\usepackage{soul}
\usepackage[normalem]{ulem}
\usepackage{fontawesome}
\DeclareUnicodeCharacter{2212}{-}
\usepackage{comment}
\usepackage{wasysym}

\Title{Probing the Dark Matter Capture Rate in a Local Population of Brown Dwarfs with IceCube Gen 2}

\TitleCitation{Probing the Dark Matter Capture Rate in a Local Population of Brown Dwarfs with IceCube Gen 2}

\Author{Pooja Bhattacharjee 
 $^{1,}$*\orcidA{} and Francesca Calore $^{2}$\orcidB{}}

\AuthorNames{Pooja Bhattacharjee and Francesca Calore}

\AuthorCitation{Bhattacharjee, P.; Calore, F.}
\address{%
$^{1}$ \quad Laboratoire d'Annecy De Physique Des Particules, Annecy 74940, France
\\
$^{2}$ \quad Laboratoire d'Annecy-le-Vieux de Physique Théorique, Annecy 74940, France; francesca.calore@lapth.cnrs.fr}

\corres{Correspondence: pooja.bhattacharjee@lapp.in2p3.fr}
\firstnote{LAPTH-061/23
}

\abstract{This study explores the potential for dark matter annihilation within brown dwarfs, investigating an unconventional mechanism for neutrino production. Motivated by the efficient accumulation of dark matter particles in brown dwarfs through scattering interactions, we focus on a mass range above 10 GeV, considering dark matter annihilation channels $\chi \chi \rightarrow \nu \bar{\nu} \nu \bar{\nu}$ through long-lived mediators. Using the projected sensitivity of IceCube Generation 2, we assess the detection capability of the local population of brown dwarfs within 20 pc and exclude dark matter-nucleon scattering with cross-sections as low as a few multiples of $10^{-36}~\rm cm^{2}$.} 
\keyword{brown dwarf; dark matter; capture rate and self-annihilation, neutrino, IceCube Gen 2} 

\begin{document}


\section{Introduction}

The brown dwarf (BD) stands as a distinctive category of celestial bodies, bridging the gap between the least massive main-sequence stars and enormous gas giants. In~contrast to typical main-sequence stars, their mass falls short of the threshold (approximately \mbox{0.07--0.08 $M_\odot$}) required for hydrogen fusion through thermonuclear processes. However, they possess the ability to sustain the burning of the limited deuterium quantities they contain. Observation suggests that they share a similar radius to that of Jupiter while weighing anywhere from 15 to 80 times the mass of Jupiter, which equals \mbox{$M_{\jupiter}\simeq 1.9\times 10^{27}$~kg}, allowing for the ignition of deuterium but not hydrogen fusion. The~concept of their existence was initially proposed in the 1960s~\cite{Kumar:1963a,Kumar:1963b,Hayashi:1963}, but~it took nearly three decades for observational confirmation to follow~\cite{Nakajima:1995,Rebolo:1995}. Owing to their low luminosity and predominant emission in the infrared spectrum, only a few thousand BDs have been identified, despite the possibility of the Milky Way hosting a billion or more of these objects. Strikingly, even though they do not exhibit particularly conspicuous characteristics, BDs have exhibited non-thermal emissions in both radio~\cite{Berger:2001rf} and X-ray wavelengths~\cite{Rutledge:2000nu}. This phenomenon seems to be primarily, if~not exclusively, associated with flares (similar to analogous occurrences in the relatively unassuming red dwarfs), and~there is speculation that it might be linked to sub-surface magnetic~activity.

 In general, BDs may not be the most suitable candidates for traditional neutrino and gamma ray production, as~they exhibit minimal thermonuclear fusion activity. However, they hold promise as potential sources for signals from dark matter (DM). If~the universe's DM consists of particles ($\chi$) with non-negligible couplings to the Standard Model (SM), compact astrophysical objects like BDs could efficiently accumulate them through scattering interactions. Recent studies have expanded their focus to include exoplanets and BDs in this context~\cite{Leane:2020wob, Leane:2021ihh, Bhattacharjee:2022lts}. The~``evaporation mass'', dependent on the temperature and density of the celestial object, defines the lightest DM particle that can be efficiently captured. BDs in particular offer advantages over the Sun for capturing sub-GeV DM due to their lower characteristic mass and temperature~\cite{Garani:2021feo}, making them prime candidates for DM capture~\cite{Leane:2020wob}.
While BDs may not be as efficient as neutron stars (NSs) in capturing DM with weak interactions, their larger size can help compensate for their lower gravitational pull. Furthermore, BDs are more numerous and closer to us compared to NSs, making them appealing targets for light and strongly interacting DM~\cite{Haberl:2006xe, Bhattacharjee:2022lts}.

The question now arises: how can we detect the consequences of DM capture in BDs? In our previous investigation~\cite{Bhattacharjee:2022lts}, our focus was on a mediator ($\phi$) that is both light and long-lived. In~this scenario, DM annihilation leads to the mediator, which subsequently decays into gamma rays, neutrinos, or any other standard model (SM) candidates. Importantly, the~mediator has a considerable lifetime~\cite{Martin:1997ns, Holdom:1985ag, Holdom:1986eq, Okun:1982xi, Kobzarev:1966qya, Dasgupta:2020dik, Nguyen:2022zwb}, resulting in these processes occurring outside the BD rather than within. Interestingly, such scenarios, which are by no means rare in theoretical considerations, are actively explored in current endeavors to discover light DM particles at particle colliders~\cite{Abdallah:2015ter}.

 In this study, we extend our exploration beyond gamma rays to include neutrinos in the search for DM. While gamma ray telescopes often provide stronger limits in the quest to detect DM annihilation, neutrino telescopes offer advantages, such as less reliance on uncertain DM density profiles and longer observation times. Neutrino telescopes can also simultaneously monitor multiple sources, making them a competitive avenue for DM research~\cite{IceCube:2021xzo, Miranda:2022kzs}.  The~gamma rays from BDs have been considered in some recent studies~\cite{Leane:2020wob, Leane:2021ihh, Bhattacharjee:2022lts, Bose:2022ola, Klangburam:2023icw}, but neutrinos have not been probed earlier from BDs. Neutrino searches offer unique advantages in the quest for DM. Neutrinos, being electrically neutral and weakly interacting, travel directly and unhindered from their sources without deflection or attenuation. This makes neutrinos valuable for gaining insights into dense sources, even those at cosmological distances, where other SM particles cannot reach. Another compelling reason to investigate neutrinos is that many SM particles ultimately decay, yielding neutrinos and gamma rays as final products. Detecting neutrinos complements the exciting recent advancements in DM annihilation searches through gamma rays~\cite{Ackermann:2011wa, Abdo:2010nc, Abramowski:2010aa, Abdo:2010dk, Aleksic:2011jx}.

 The IceCube Neutrino Observatory, located at the South Pole, is one of the leading scientific facilities designed to detect high-energy neutrinos in the TeV--PeV energy range~\cite{IceCube:2019pna}. It consists of a cubic kilometer array of optical sensors embedded deep in the polar ice. IceCube's primary mission is to investigate the origins and characteristics of the most energetic neutrinos in the universe, shedding light on extreme astrophysical events. Its innovative design and remote location shield it from cosmic rays, allowing for precise neutrino detection. IceCube has made significant contributions to astrophysics and particle physics, uncovering insights into neutrinos and high-energy cosmic phenomena~\cite{IceCube:2018ndw, IceCube:2021xar, IceCube:2021uhz} and has been engaged in the quest to identify high-energy neutrino emissions from specific time-integrated point sources, encompassing comprehensive all-sky investigations~\cite{IceCube:2021xar, IceCube:2021uhz, IceCube:2018ndw}.

The next planned extension of IceCube, IceCube Gen 2, plans to probe the high-energy neutrino spectrum from TeV to PeV energies~\cite{IceCube:2019pna, IceCube-Gen2:2023vtj}, offering five times better sensitivity than the current IceCube detector. It will include 120 new strings with larger spacing, enhancing sensitivity for neutrinos above 10 TeV and targeting detections over 100 PeV~\cite{IceCube-Gen2:2023vtj}, and will improve the detection thresholds down to 1 GeV. This will be very promising, especially for detecting the fainter sources~\cite{IceCube:2021oqo}, and will make IceCube Gen 2 the leading neutrino observatory to support diverse scientific avenues such as DM searches, particle physics, and~supernova neutrino detection~\cite{Ishihara:2019aao}, which are yet to be addressed by current IceCube~generation.


In our earlier study~\cite{Bhattacharjee:2022lts}, we showed that the BD can be sensitive to both spin-dependent and spin-independent DM-proton scattering due to its hydrogen content. DM particles from the galactic halo can elastically interact with nuclei, resulting in their capture and thermalization in the BD.  In~this study, we search for the cumulative emission of neutrino signals from the population of local BDs within 20 pc with the projected sensitivity of IceCube Gen 2 to investigate the DM capture rates in BDs. We also perform a comparative study between the projected sensitivity of IceCube Gen 2 and the sky survey of astrophysical neutrinos performed by IceCube using 10 years of muon track events~\cite{IceCube:2021xar, IceCube:2019cia}.

The remainder of the paper is arranged as follows. In Section~\ref{sec:formulation}, we describe the formulation for the DM capture rate in BDs, their annihilation into long-lived mediators, and~the expected spectra for neutrino flux. 
In Section~\ref{source}, we briefly review the sources we considered for this study. 
In Section~\ref{results}, we present our results followed by a methodology to utilize the IceCube and IceCube Gen 2 sensitivity for our sources. Finally, we conclude our study in Section~\ref{conclusion}.
\section{DM Capture and Annihilation in~BDs}
\label{sec:formulation}
In this section, we briefly review the mechanism of DM capture by BDs and their self-annihilation to neutrinos. 
We divide the formulation part into two steps:
(i) the capture of DM into BDs, which depends notably on the local density and velocity distribution; and 
(ii)~the~DM self-annihilation to a low-mass scalar mediator that will eventually decay to four neutrinos outside~BD.

\subsection{DM Capture~Rate}
\label{sec:dm_capture}

When DM particles traverse celestial objects and undergo one or multiple collisions, they can lose enough energy to become captured. The~rate at which DM is captured is directly proportional to the DM density. In~our investigation, we characterize the DM distribution in the galactic halo using the cuspy Navarro--Frenk--White (NFW) density profile~\cite{Navarro:1996gj}, as~expressed by:
\begin{equation}
 \rho_{\chi \rm{NFW}} = \frac{\rho_{0}}{(r/r_{s})^{\gamma} (1+(r/r_{s}))^{3-\gamma}}\,\, .
 \label{eqn:dm_density_nfw}
\end{equation}

 Here, 
 $r_{s}$ is the scale radius for the NFW profile, and~$\rho_{0}$ is normalized to the local DM density value. The~parameter $\gamma$ determines the inner slope of the DM profile. We adopt the values of $r_{s}$ and $\rho_{0}$ from Ref.~\cite{Calore:2022stf}. 

In our calculations, we account for both single and multiple scattering of DM candidates following the formulation of Ref.~\cite{Bramante:2017xlb, Ilie:2020}. The~probability of DM particles scattering $N$ times before being captured is given by:
\begin{equation}
p_{\rm N}(\tau)=2\,\int_{0}^{1}\, dy \,\frac{y\,e^{-y\tau}\,(y\tau)^N}{N!}\,\, .
\label{eqn:p_N}
\end{equation}

Here, $\tau$ is the optical depth, which depends on saturation cross-section ($\sigma_{\rm sat}$), i.e.,~$\tau$=$\frac{3}{2} \frac{\sigma_{\chi n}}{\sigma_{\rm sat}}$ of the compact object with mass $M_{\star}$ and radius $R_{\star}$, while $\sigma_{\chi n}$ is the DM-nucleon scattering cross-section, and $\sigma_{\rm sat}$ = $\pi~R_\star^{2}/N_{n}$, where $N_{n}=M_\star/m_n$ is the number of nucleons in the target, with~$m_{n}$ being the nucleon~mass.

 The total capture rate ($C_{\rm tot}$) after single and multiple scattering is expressed as:
\begin{equation}
 C_{\rm tot}(r) = \sum_{n=1}^{\infty} C_{\rm N}(r) \, ,
 \label{eqn:c_tot}
\end{equation}
where the capture rate associated with $N$ scatterings, or $C_{\rm N}$ in a DM environment with number density $n_{\chi}=\rho_\chi/m_\chi$, is given by:

\begin{adjustwidth}{-\extralength}{0cm}
\centering 
\begin{equation}
C_{\rm N}(r) = \frac{\pi\,R_\star^{2}p_{N}(\tau)}{1-2GM_\star/R_\star} \frac{\sqrt{6}n_{\chi}(r)}{3\sqrt{\pi}\bar{v}(r)} \times \left[(2\bar{v}^{2}+3v_{\rm esc}^{2})-(2\bar{v}(r)^{2}+3v_{N}^{2})\exp\left(-\frac{3(v_{\rm N}^{2}-v_{\rm esc}^{2})}{2\bar{v}(r)^{2}}\right)\right]\,\, 
\label{eqn:c_N}
\end{equation}
\end{adjustwidth}
with \vspace{-6pt}
\begin{align}
v_{\mathrm{N}} &= v_{\mathrm{esc}}(1-\beta_{+}/2)^{-N/2},  	\nonumber \\
\beta_{+} &= \frac{4m_{\chi}m_{n}}{(m_{\chi}+m_{n})^{2}},	\nonumber
\end{align}
where $r$ is the galactocentric distance, and the other terms are specified as: (i) escape velocity ($v_{\rm esc}$): $v_{\rm esc}$ = $\sqrt{2G M_{\star}/R_{\star}}$, where G is the gravitational constant; (ii) 'typical' velocity in the DM rest frame ($v_{0}$): $v_{0} = \sqrt{8/(3\pi)}\bar{v}$, where $\bar{v}$ is root mean square DM velocity; note that $\bar{v}$ is related to the circular velocity at distance $r$, $v_{\rm c}(r)$, as~$\bar{v}(r) = 3/2 v_{\rm c}(r)$; {and (iii) circular velocity ($v_{\rm c}$(r)): $v_{\rm c}(r) = \sqrt{\frac{G M(r)}{r}}$, where $M(r)$ is the total Galactic mass enclosed within $r$~\cite{Sofue:2013}.}

\subsection{DM Annihilation~Rate}
\label{sec:dm_annihilation}

In Section~\ref{sec:dm_capture}, we discussed the formalism of DM capture in the BD. In~this section, we explore the annihilation of the captured DM particles. The~time derivation of the number of trapped DM particles $N(t)$ inside BD is as follows:
\begin{equation}
\frac{{\rm d}N(t)}{{\rm d}t} = C_{\rm tot} - C_{\rm ann}N(t)^{2}\,- C_{\rm evap}N(t)\, .
\label{eqn:N_variation}
\end{equation}

Here, $C_{\rm tot}$ is the total capture rate of DM obtained in Equation~(\ref{eqn:c_tot}), and~$C_{\rm ann}$ and $C_{\rm evap}$ are the thermally averaged annihilation cross-section over the containment volume $V_{\rm c}$ and the evaporation rate of DM, respectively. The~recent study by Ref.~\cite{Garani:2021feo} estimated the evaporation mass of BD around $\sim$0.7 GeV. Thus, we can safely neglect the effect of $C_{\rm evap}$~\cite{Garani:2021feo} for the DM mass range of our~interest. 

Equation~(\ref{eqn:N_variation}) admits the solution
\begin{equation}
N(t) = C_{\rm tot}\,t_{\rm eq} \tanh (t/t_{\rm eq}+t_{0}/t_{\rm eq})
\label{eqn:Neq_time}\, ,
\end{equation}
 where $t_{\rm 0}$ is an integration constant related to N(t=0) 
  and
\begin{equation}
t_{\rm eq} \equiv (C_{\rm ann}C_{\rm tot})^{-1/2}\, .
\label{eqn:equilibrium_time}
\end{equation}

In Equation~(\ref{eqn:equilibrium_time}), $t_{\rm eq}$ is the equilibrium timescale over which the equilibrium between $C_{\rm tot}$ and $C_{\rm ann}$ will take place. Once equilibrium is reached, the~DM annihilation rate, $\Gamma_{\rm ann}$, can be written as
\begin{equation}
\Gamma_{\rm ann}(t) =\frac{C_{\rm ann}N(t)^2}{2}\to \frac{C_{\rm tot}}{2}\,\, .
\label{eqn:cap_ann}
\end{equation}

 In Equation~(\ref{eqn:cap_ann}), the~factor of 2 derives from the fact that two DM particles participate in each~self-annihilation.

\subsection{Neutrino~Spectrum}
\label{spectrum}

In this paper, we delve into the realm of particle physics to determine the neutrino flux resulting from DM annihilation. Generally, when DM particles directly annihilate into \mbox{2-body} final states, such as (i) $\chi \chi \rightarrow \mu^+ \mu^-$, (ii) $\chi \chi \rightarrow \nu \overline{\nu}$, (iii) $\chi \chi \rightarrow~\tau^+ \tau^{-}$, \mbox{(iv)~$\chi \chi \rightarrow W^+ W^-$},  (v) $\chi \chi \rightarrow b\bar{b}$, and so on, they can generate a detectable neutrino signal, as~indicated in Ref.~\cite{Dasgupta:2012bd}. These scenarios are of interest due to their potential for sensitivity in measuring the $\sigma_{\chi n}$ parameter using neutrino~telescopes.

 In the case of DM annihilation within celestial bodies like BDs, there is a possibility that if DM particles directly annihilate into SM states, they may become trapped within the BD, leading to its heating. However, the~likelihood of detecting a neutrino signal increases when DM annihilation to SM states occurs through long-lived mediators that can escape the BD's surface, as~suggested by Refs.~\cite{Pospelov:2007mp, Pospelov:2008jd, Batell:2009zp, Dedes:2009bk, Fortes:2015qka, Okawa:2016wrr, Yamamoto:2017ypv, Holdom:1985ag, Holdom:1986eq, Chen:2009ab, Rothstein:2009pm, Berlin:2016gtr, Cirelli:2016rnw, Cirelli:2018iax}. The~decay products of these mediators may yield observable signals in current or future neutrino~telescopes.

 Nonetheless, these mediators could also interact with SM constituents within the BD before escaping, potentially leading to a significant reduction in the observed neutrino flux. As~demonstrated in~\cite{Leane:2021ihh, Bhattacharjee:2022lts}, by~selecting appropriate model parameters, it is feasible to reduce this attenuation considerably. Therefore, we assume that all mediator particles ultimately decay outside the BDs, producing detectable neutrino flux. The concept of captured DM annihilation through long-lived mediators has been explored previously in the context of the Sun~\cite{Silk:1985ax, Krauss:1985aaa, Jungman:1994jr, Pospelov:2008jd, Batell:2009zp, Bell:2011sn, IceCube:2012ugg, Baratella:2013fya, Danninger:2014xza, Smolinsky:2017fvb, Leane:2017vag, HAWC:2018szf, Nisa:2019mpb, Niblaeus:2019gjk, Xu:2020lrn, Bell:2021pyy, Leane:2021tjj, Xu:2021glr, Bell:2021esh} and~more recently in the context of exoplanets~\cite{Leane:2021tjj, Leane:2021ihh, Bhattacharjee:2022lts}.

 We make an assumption that DM annihilates into mediators that are both ``light'' and ``long-lived'', described as $\chi\chi\to \phi\phi$. Here, ``long-lived'' implies that the decay range, denoted as $L$, greatly exceeds the characteristic scale $R_\star$, while ``light'' signifies that we operate under the simplifying assumption that the mediator's mass $m_{\phi}$ is much smaller than that of the DM particle $m_{\chi}$. 

This choice is motivated by secluded DM models~\cite{Pospelov:2007mp, ArkaniHamed:2008qn}, wherein DM annihilates into two light vector bosons or a similar mediator, each of which subsequently decays into SM particles and can be experimentally observed~\cite{Pospelov:2008jd, Rothstein:2009pm, Bell:2011sn}.  Although~our primary focus centers on light and long-lived mediators, our approximation remains valid within a finite range of mediator mass $m_{\phi}$. Assuming our stated hypothesis, the~neutrino spectrum is envisaged as a box-shaped distribution, as~expressed~\cite{Ibarra:2012dw} in Equation~(\ref{eq:box_function}):
\begin{equation}
\frac{dN_{\nu}}{dE_{\nu}}=\dfrac{4}{\Delta E}\Theta(E_{\nu}-E_-)\Theta(E_+-E_{\nu})\, ,
\label{eq:box_function}
\end{equation}
 where $\Theta$ represents the Heaviside-theta function. The~upper and lower bounds of neutrino energy are denoted by $E_{\pm}=(m_{\chi}\pm \sqrt{m_{\chi}^2-m_{\phi}^2})/2$, and~the width of the box function is defined as $\Delta E=\sqrt{m_{\chi}^2-m_{\phi}^2}$.

 In this study, our objective is to assess the sensitivity of neutrino telescopes to $\sigma_{\chi n}$ for DM annihilation into neutrinos. Specifically, we plan to optimize the sky survey sensitivity of IceCube Gen 2~\cite{IceCube:2019pna} (and also for IceCube, utilizing 10 years of data~\cite{IceCube:2021xar, IceCube:2019cia}), with~a focus on the muon neutrino flux originating from DM capture in~BDs.

 The expression of the differential neutrino flux reaching Earth due to captured DM annihilation through a long-lived mediator is:
\begin{eqnarray}
E_{\nu}^2 \, \frac{d \phi_{\nu}}{d E_{\nu}} &=& \frac{\Gamma_{\rm ann}}{4 \, \pi \, d_\star^{2}} \times \left( \frac{1}{3} E_{\nu}^2 \frac{d N_{\nu}}{d E_{\nu}} \right)  \times  \left( e^{-\frac{R_\star}{\eta c \tau}} - e^{-\frac{d_\star}{\eta c \tau}} \right)\, ,
\label{eqn:dm_flux}
\end{eqnarray}
 where $d_\star$ represents the Earth's distance from the BD location. This formula also includes the contribution of the ``survival'' probability ($P_{\rm surv}=e^{-\frac{R_\star}{\eta c \tau}} - e^{-\frac{d_\star}{\eta c \tau}}$) of neutrinos reaching Earth's detectors, which is near unity 
  within the range of mediator decay lengths. Achieving this entails either a sufficiently long mediator lifetime ($\tau_{\phi}$) or a considerably high boost, denoted as $\eta = m_{\chi}/m_{\phi}$. Under~these conditions, the~mediator's decay length $L \approx \eta c \tau_{\phi}$ exceeds $R_\star$. 
\section{Source~Selection}
\label{source}

Since the first discovery of BD, several hundred BDs have been identified, mostly in the large-scale optical and near-infrared imaging surveys performed by the Two-Micron All-Sky Survey (2MASS,~\cite{Skrutskie:2006}) and the Sloan Digital Sky Survey (SDSS,~\cite{york:2000}). In~the recent era, large-area surveys like the UKIRT Infrared Deep Sky Survey (UKIDSS,~\cite{Lawrence:2007}), the~Canada--France Brown Dwarfs Survey (CFBDS,~\cite{Delorme:2008}), etc., used much deeper imaging technology. As~a result, the~detection of the fainter and cool brown dwarfs has increased.

For this study, we select the same set of BDs that we considered for our previous study~\cite{Bhattacharjee:2022lts} and follow the same nomenclature. For~the details, please see Section~2 and Table~1 of Ref.~\cite{Bhattacharjee:2022lts}. Apart from considering those closest BDs, in~this study, we also estimate the distribution of local BDs using the list of detected BDs mentioned in Ref.~\cite{BDlist}. We only use the $\sim$800 T and Y type BD sources ($N_{\rm total}$ lie within 1200 pc in Equation~(\ref{eqn:BD_N_tot})) from Ref.~\cite{BDlist} and derive the local population density of detected BDs ($n_{\rm BD_{\rm loc}} (b,l,D)$) as a function of latitude ($b$), longitude ($l$), and distance ($D$) in Equation~(\ref{eqn:BD_n_loc}). We find that the distribution with longitude is almost constant, whereas with~latitude, it follows a normal distribution with a mean of 5.69 and a sigma of 30.49. The~distribution with distance almost varies the same as the $\rm chi^2$ 
 distribution (with degrees of freedom = 2.58). This population study has an advantage in predicting an improvement in scattering cross-section bounds with respect to individual local BDs, as~we discuss in Section~\ref{results}.
\begin{equation}
N_{\rm total} = \int_{\rm bmin}^{\rm bmax}\,\int_{\rm lmin}^{\rm lmax}\, \int_{\rm dmin}^{\rm dmax}\, n_{\rm BD}(b,l,D)\, \sin b\, db\, dl\, D^2\, d\it{D},
 \label{eqn:BD_N_tot}
\end{equation}
where $b_{\rm min}~=~-90^{\circ}$, $b_{\rm max}~=~+90^{\circ}$, $l_{\rm min}~=~-180^{\circ}$, $l_{\rm max}~=~+180^{\circ}$, $D_{\rm min}$ = 0 pc, and $D_{\rm max}$ = 1200 pc.
\begin{equation}
n_{\rm BD_{\rm loc}} (b,l,D) = K~\times~P(b)~\times~G(l)~\times~H(D),
 \label{eqn:BD_n_loc}
\end{equation}
where $K$ is a normalization constant to guarantee the correct integral of Equation~(\ref{eqn:BD_N_tot}); $P(b)$, $G(l)$, and $H(D)$ are the probability distribution functions (PDFs) of latitude ($b$), longitude ($l$), and distance ($D$), respectively, normalized to unity. These PDFs are derived from the observed BD sample~\cite{BDlist}. 

Following this distribution, there will be around $\sim$30 BDs within 20 pc that we will use later for our~study.

\section{Results}
\label{results}

The sensitivity curve of a telescope is pivotal for distinguishing the threshold signal from background noise. In~this study, our focus is on neutrinos generated through the annihilation of captured DM in BDs, as~observed by the neutrino telescopes from the south pole 
 IceCube and IceCube Gen~2 
. Neutrino interactions at IceCube and IceCube Gen 2 generally take on two topologies: track-like and cascade-like. Track-like signals primarily arise from charged-current interactions involving muon (anti-)neutrinos ($\nu_{\mu}$ and $\bar{\nu}{_{\mu}}$) with nucleons, resulting in the production of high-energy muons and anti-muons. Tau (anti-)neutrinos ($\nu_{\tau}$ and $\bar{\nu}{_{\tau}}$) can also generate energetic muons through charged-current interactions, but the branching ratio is not large enough to provide a large contribution. On~the other hand, cascade-like events mainly emerge from charged-current interactions involving astrophysical electrons or tau (anti-)neutrinos, as~well as neutral current interactions of any neutrino type. These cascade-like events are generally less suitable for point-source studies. Furthermore, the~abundance of track-like events greatly surpasses that of cascade-like events, primarily because neutrinos can interact far outside the detector before IceCube detects the secondary muon~\cite{IceCube:2021xar}.

This paper is dedicated to exploring track events collected or to be probed by IceCube Gen 2, with~a focus on the significance of astrophysical signals from point-like sources. Our analysis considers recent all-sky searches for point-like neutrino sources~\cite{IceCube:2021xar} and the time-integrated analysis~\cite{IceCube:2019cia} conducted by IceCube using data spanning 10 years (6 April 2008 to~8 July 2018). IceCube measurements have confirmed the presence of an astrophysical neutrino flux across various energy ranges and neutrino types. The~expanded coverage of IceCube Gen 2 enhances detection capabilities, enabling the detection of even fainter fluxes from individual sources, surpassing current limitations. For~IceCube Gen 2, we examine the discovery potential for a single point-like source for a period of 10 years with 300~s of exposure time~\cite{Ye:2022vbk, IceCube-Gen2:2021tmd}. Our analysis focuses on track events predominantly originating from muon (anti-)neutrinos ($\nu_{\mu}$ and $\bar{\nu}{_{\mu}}$), which traverse the detector from all directions, as~well as neutrino track events beginning within the instrumented~volume.

\begin{figure}[H]
\includegraphics[width=10 cm]{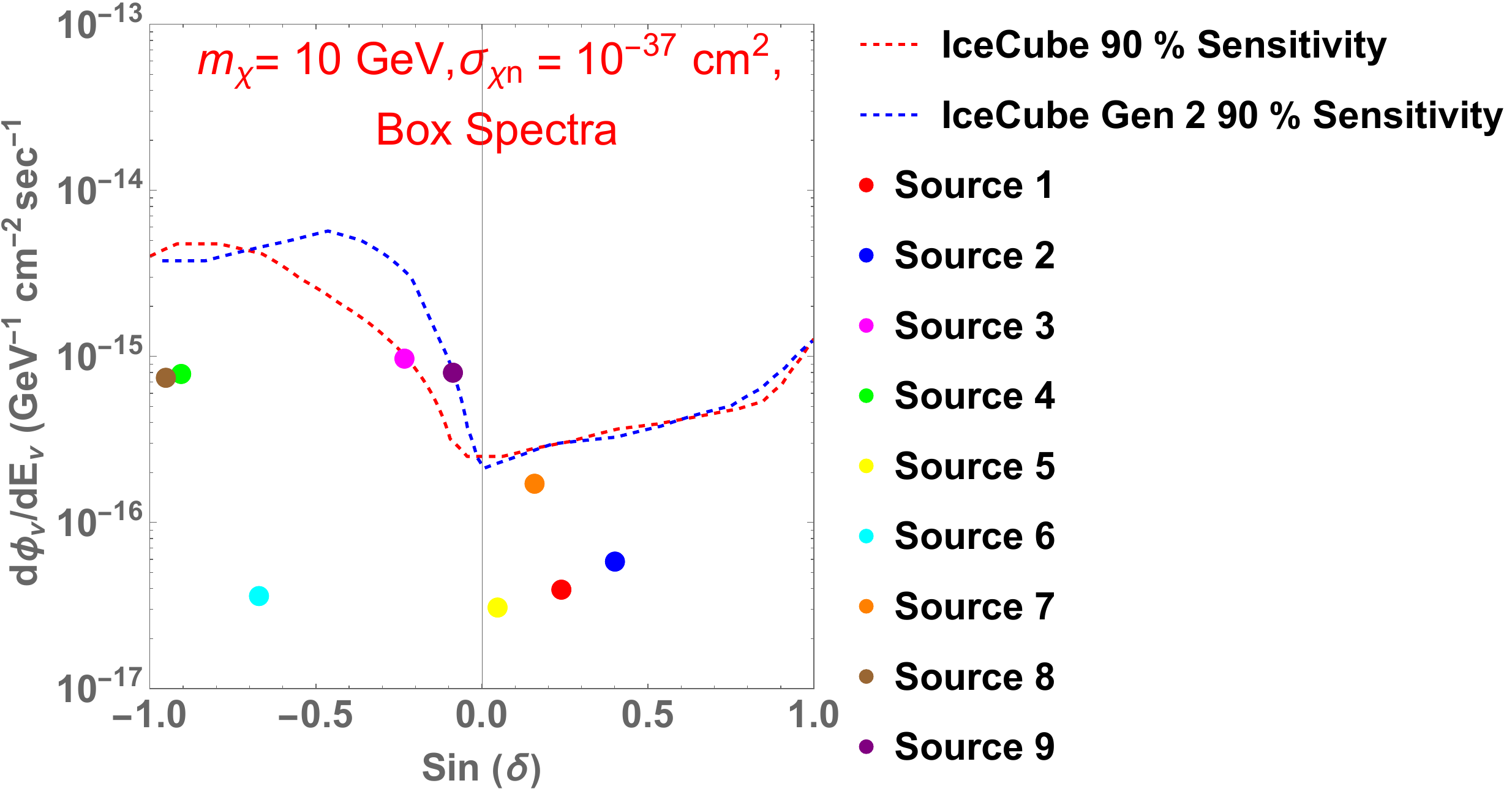}
\caption{Differential 
 flux as the function of the sine 
  of the declination angle $(\delta)$ for IceCube and IceCube Gen 2, where the points define the individual differential neutrino flux for each BD obtained from Equation~(\ref{eqn:dm_flux}) for DM mass, $m_\chi$ = 10 GeV, and~scattering cross-section, $\sigma_{\chi n}$ = $10^{-37}$~cm$^{-2}$, from~box-like spectra. The~red and blue dashed lines denote the 90\% C.L. sensitivity of IceCube~\cite{IceCube:2019cia} and IceCube Gen 2~\cite{IceCube-Gen2:2021tmd}, respectively.} 
\label{fig2}
\end{figure}

 Figure~\ref{fig2} provides insight into the 90\% confidence level (C.L.) sensitivity of both IceCube~\cite{IceCube:2019cia} and IceCube Gen~2~\cite{IceCube-Gen2:2021tmd} by adopting a source spectrum with a differential flux $\frac{dN}{dE}\propto\rm E^{-2}$ as a function of source declination (sin $\delta$). The~`points' denote the individual differential muon neutrino flux obtained from Equation~(\ref{eqn:dm_flux}) for each BD for DM mass, \mbox{$m_\chi$ = 10 GeV,} and~scattering cross-section, $\sigma_{\chi n}$ = $10^{-37}~\rm cm^{2}$, from~box-like spectra (by assuming annihilation through a long-lived mediator). 
 
 The~details of our selected source are available in our earlier study~\cite{Bhattacharjee:2022lts}. (Source 
 1---sin $\delta=$ 0.238714; Source 2---sin $\delta=$ 0.39945; Source 3---sin $\delta=$ $-$0.231999; Source 4---sin $\delta=$ $-$0.905029; Source 5---sin $\delta=$ 0.0452918; Source 6---sin $\delta=$ $-$0.671652; Source 7---sin $\delta=$ 0.159834; Source 8---sin $\delta=$ $-$0.950165; Source 9---sin $\delta=$ $-$0.0875628.) From~Figure~\ref{fig2}, we can also conclude that, for IceCube and IceCube Gen 2, the highest sensitivity is observed around the equator, aligning with the optimal discovery potential, while it is the weakest for declination near sin $\delta =~\pm 1$, signifying that detecting events of interest amidst the background requires a stronger signal in these regions. Thus, with~the current sensitivity of IceCube and IceCube Gen 2, only two of our sources, i.e.,~Source 3 (sin $\delta = -0.232$) and Source 9 (sin $\delta = -0.087$), which lie in the vicinity of the 90\% C.L. sensitivity range, have the potential to be detected by IceCube and IceCube Gen 2 for certain conditions. Source 9 is especially interesting, as it lies above the 90\% C.L. sensitivity curve of IceCube and also shows the potential to be detected by IceCube Gen~2.

\begin{figure}[H]
\includegraphics[width=6.5 cm]{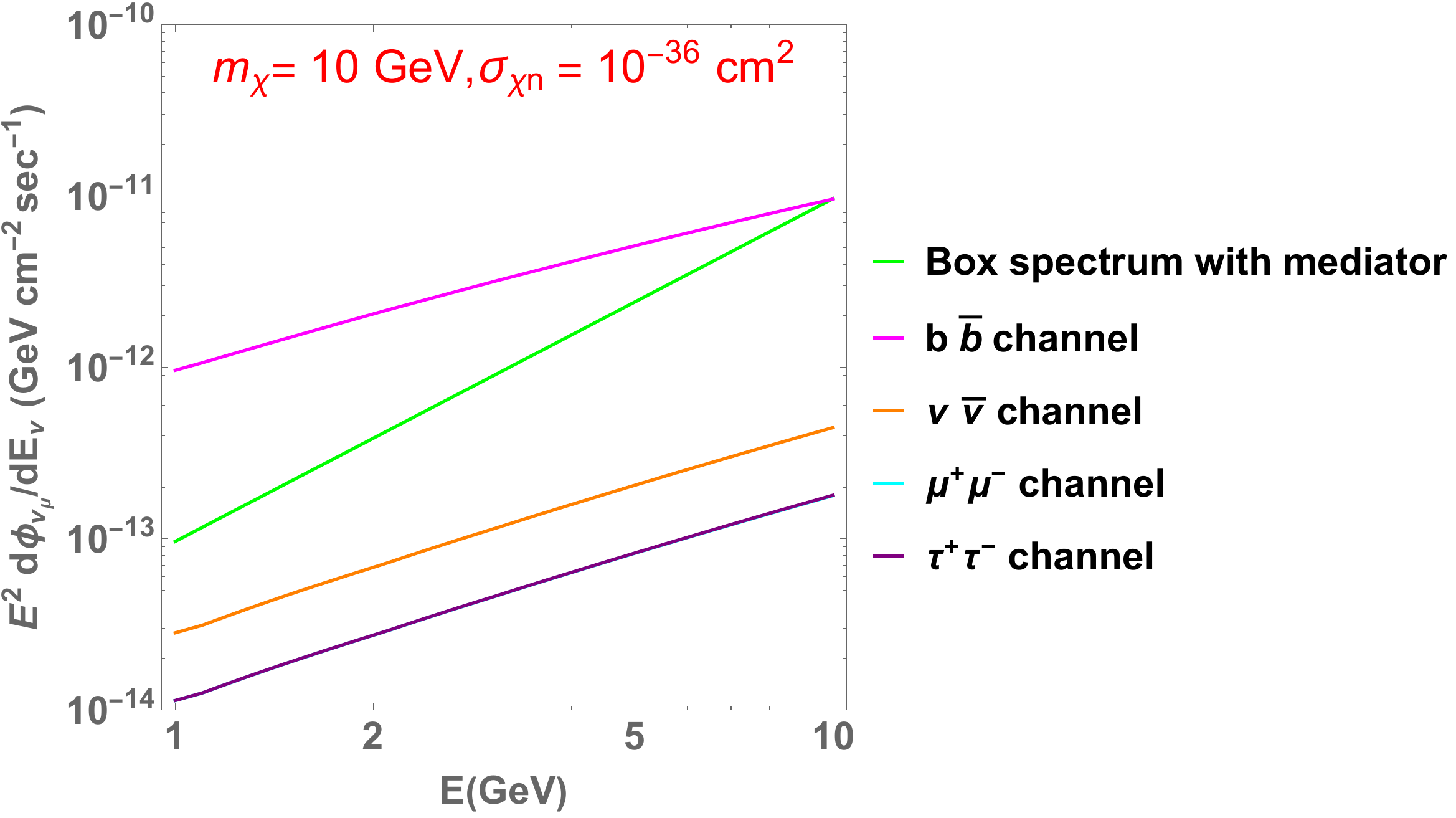}
\includegraphics[width=6.5 cm]{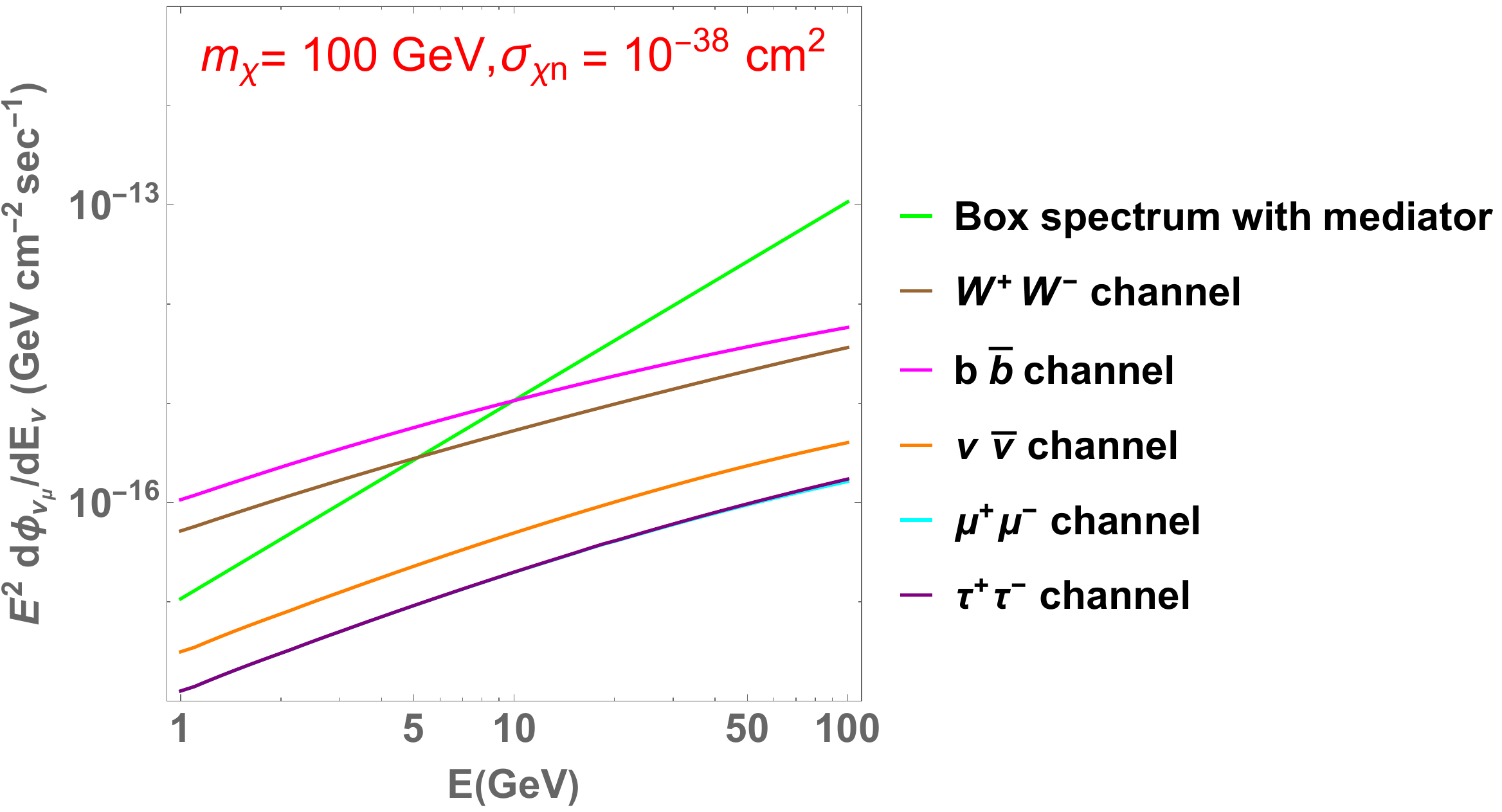}
\includegraphics[width=6.5 cm]{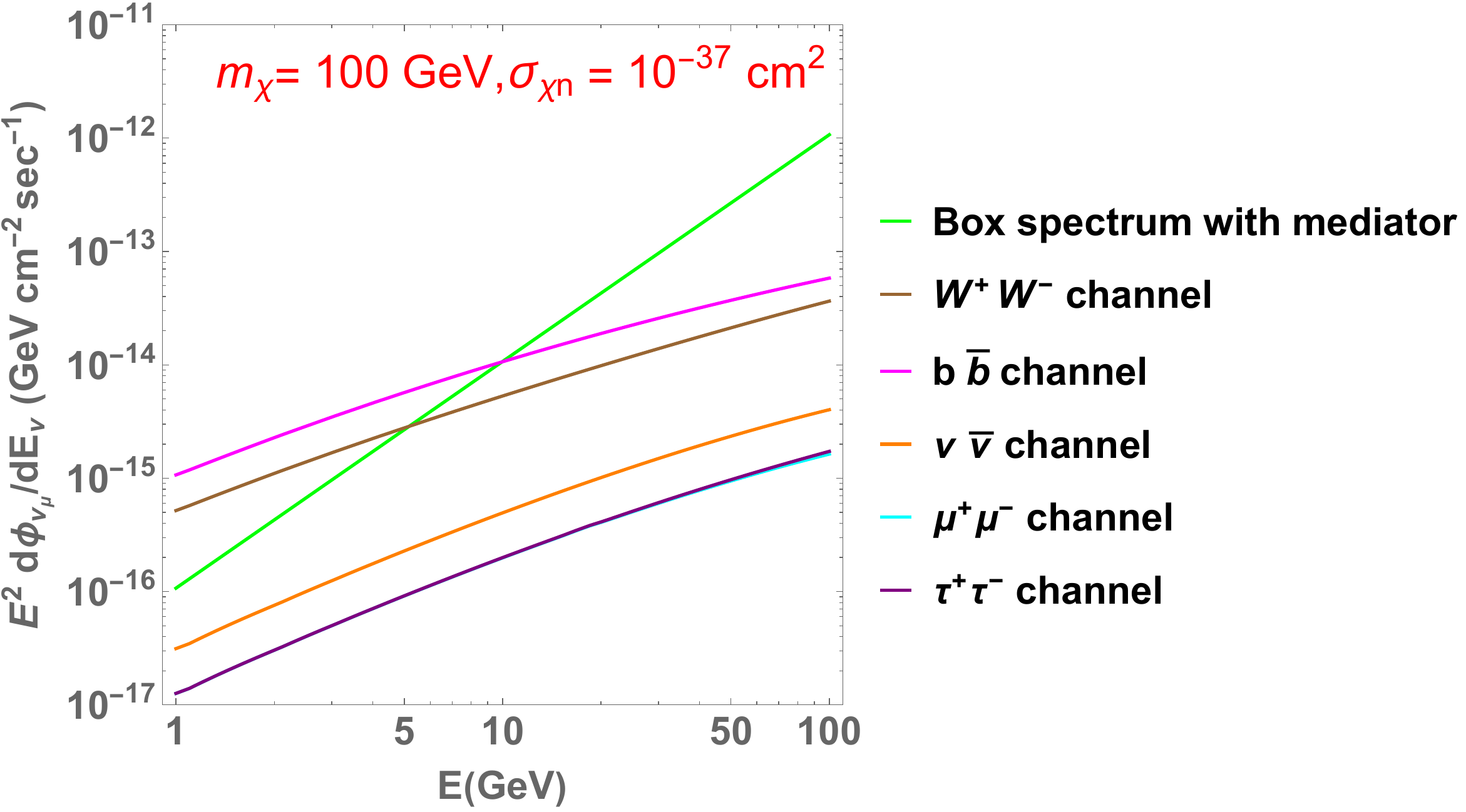}
\includegraphics[width=6.5 cm]{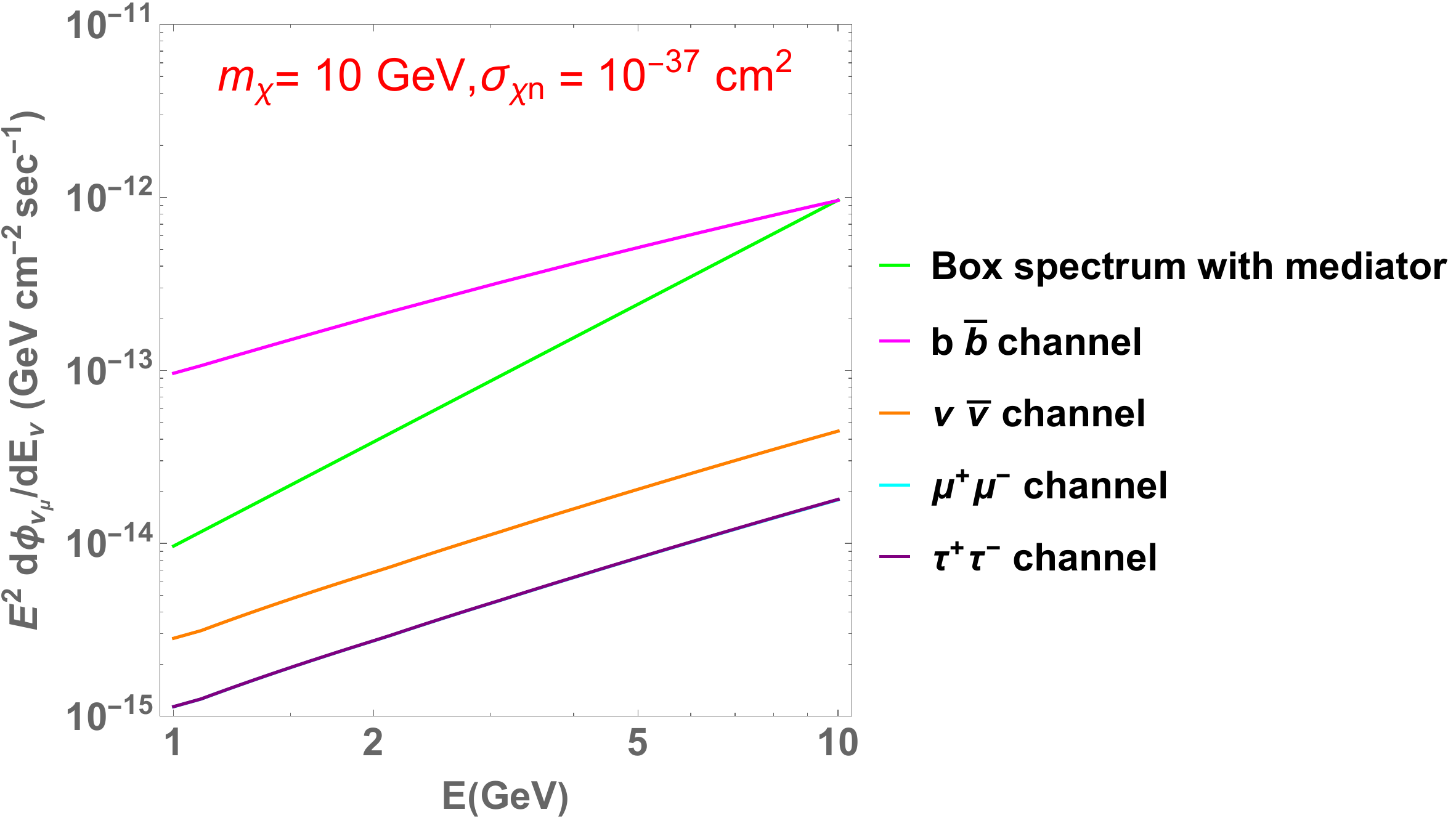}
\includegraphics[width=6.5 cm]{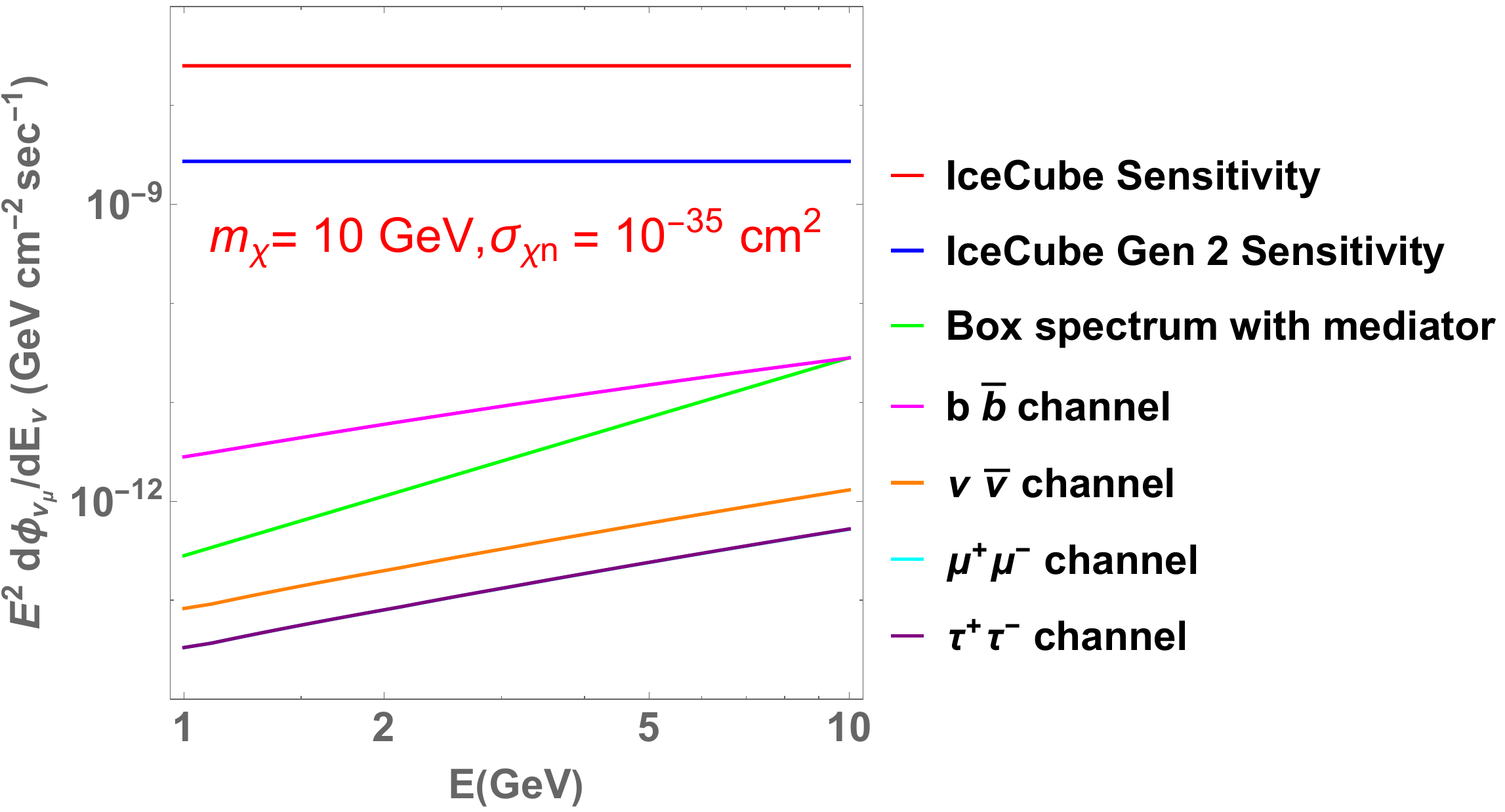}
\caption{Differential 
 neutrino flux of Source 9 for several DM masses and $\sigma_{\chi n}$ values for both box spectra and direct spectra associated with neutrinos. The~solid green lines denote the box spectra, whereas $b\bar{b}$, $\mu^+ \mu^-$, $\nu \overline{\nu}$, $\tau^+ \tau^{-}$, and~$W^+ W^-$  annihilation channels are denoted by magenta, cyan, orange, purple, and brown lines, respectively. The~solid red and blue lines define the sensitivity of IceCube and IceCube Gen~2, respectively, taken from Ref.~\cite{IceCube:2019pna}.}  
\label{fig3}
\end{figure}

IceCube and IceCube Gen 2 possess the ability to explore both spin-dependent (SD) and spin-independent (SI) DM-nucleon scattering cross-sections. For~BDs, the~constraints obtained are nearly equally robust for SD and SI cross-sections, given that BDs primarily consist of hydrogen. In~our previous study~\cite{Bhattacharjee:2022lts}, we confirmed that the age of our selected BDs aligns with their equilibrium age ($t_{eq}$). Therefore, for~this study, we solely focus on the DM-nucleon scattering cross-section for equilibrium. Our analysis, as~depicted in our earlier study~\cite{Bhattacharjee:2022lts}, underscores that BDs represent a promising avenue for probing small DM-nucleon cross-sections. In~Section~\ref{spectrum}, we delve into the advantages of opting for a box spectrum (i.e., DM annihilates via long-lived mediators) over direct spectra. For~the sake of comprehensive comparison, in~Figure~\ref{fig3}, we present differential flux limits of Source 9 for several conditions of $m_\chi$ and $\sigma_{\chi n}$ for both box spectra (green) and direct spectra associated with neutrinos. In~the case of direct spectra, our selections encompass: (i) $\mu^+ \mu^-$ (cyan), (ii) $\nu \overline{\nu}$ (orange), (iii) $\tau^+ \tau^{-}$ (purple), (iv) $W^+ W^-$ (brown), and (v) $b\bar{b}$ (magenta), with an assumed 100\% branching ratio to individual SM final states.  In~Figure~\ref{fig3}, the~solid red and blue lines define the sensitivity of IceCube and IceCube Gen 2, respectively (from Ref.~\cite{IceCube:2019pna}), for detecting the neutrino excess of 5$\sigma$ originating from a point source situated at the celestial equator ($\delta = 0^{\circ}$). 

 Figure~\ref{fig3} further signifies that the flux expected from BDs (here we choose Source 9, as it is expected to provide the best limits) is a few orders of magnitude lower than the current (and future) sensitivity of IceCube (and IceCube Gen 2). This motivates us to perform the population study of local BDs. We consider the population of local BDs and estimate whether, with~their present and expected sensitivity, IceCube and IceCube Gen 2 will be able to detect the cumulative emissions coming from the local population of BDs derived in Section~\ref{source}.\\

\begin{figure}[H]
\includegraphics[width=13.5 cm]{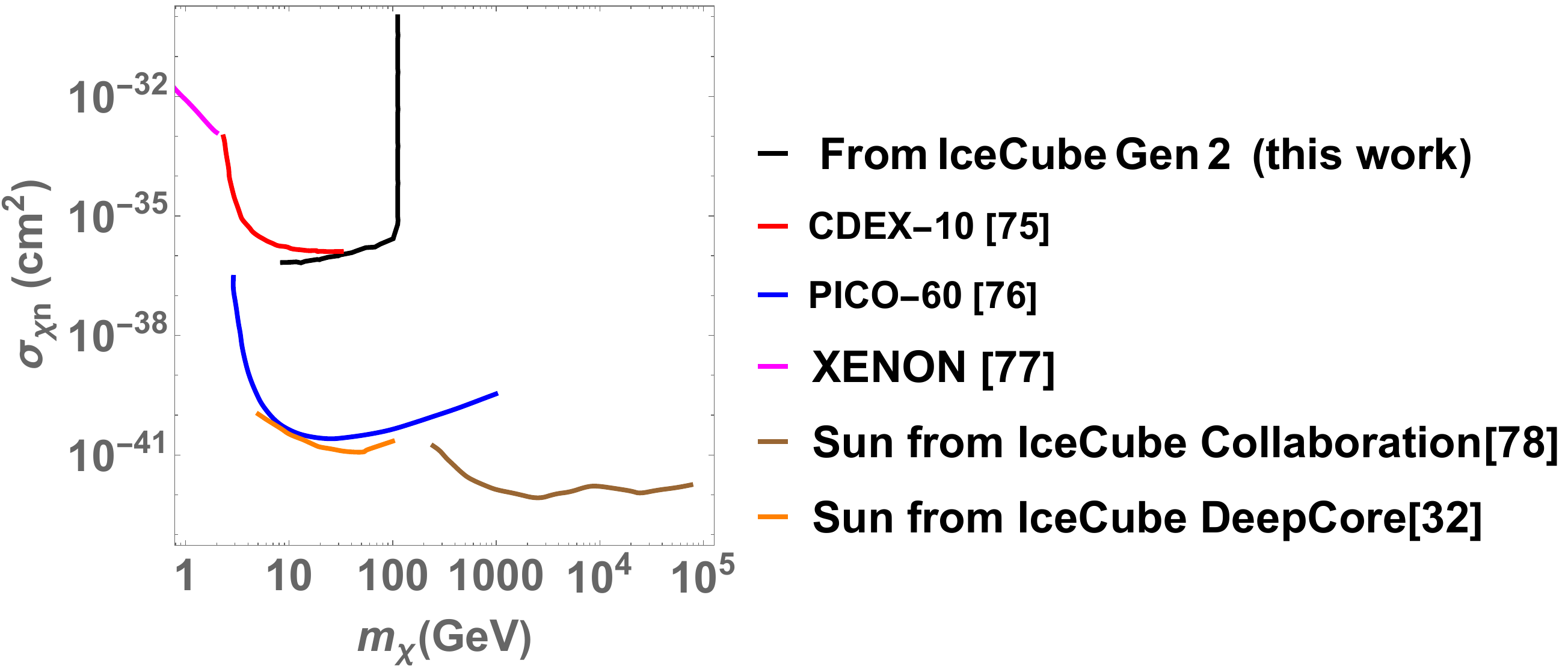} 
\caption{Comparison between the limits obtained from our work with other literature studies. We show the constraints from direct detection, such as CEDX-10~\cite{CDEX:2018lau}, PICO-60~\cite{PICO:2019vsc}, and~XENON1T~\cite{XENON:2019zpr}, on~the DM-nucleon spin-dependent cross-section, as~well as from DM annihilation from Sun from IceCube~\cite{IceCube:2021xzo,refId0}.
\label{fig6}}
\end{figure}

 \textls[-15]{In Figure~\ref{fig6}, we report the sensitivity corresponding to the cumulative emission from the expected BDs within 20 pc  using the deepest projected sensitivity of IceCube~Gen~2~\cite{IceCube:2019pna}. As~for our interest, we only focus on the nearby BDs within~20 pc from Earth and, from the distribution function of Equation~(\ref{eqn:BD_N_tot}), we predict around 30 BDs within this~distance.} 


\unskip

In Figure~\ref{fig6}, we further conduct a comparison between our results and the outcomes of astrophysical and direct DM detection experiments. It is particularly intriguing to compare our constraints with the direct detection limits established by CEDX-10~\cite{CDEX:2018lau}, PICO-60~\cite{PICO:2019vsc}, and~XENON1T~\cite{XENON:2019zpr} regarding DM-nucleon SD scattering cross-sections. We could also technically compare our bounds with the SI limits from recent studies, but~they are much stronger than our obtained limits. So, a~meaningful comparison only arises with the SD limits, where our constraints are competitive. In~our previous study~\cite{Bhattacharjee:2022lts}, we highlighted the unique advantage of the constraints derived from BDs, extending to DM masses below a few GeV compared to the existing direct detection limits. This advantage holds for neutrinos as well but, due to the current sensitivity of IceCube Gen 2, we restrict ourselves to probing the $\sigma_{\chi n}$ down to at least $\approx 10^{-36}$ cm$^{2}$ between 10--100~GeV. 

\section{Discussion and Conclusions}
\label{conclusion}

 The nature of DM remains elusive despite  multiple experimental and phenomenological efforts to unravel its nature. Indirect detection through neutrinos provides a complementary way to explore DM. IceCube has a very active program for the indirect detection of DM and can set the limits in SI/SD DM-nucleon cross-section for GeV-to-TeV parameter space. In~recent times, several studies have probed the neutrinos originating from captured DM in the core of  celestial objects~\cite{Dasgupta:2012bd, Bose:2021yhz, IceCube:2021xzo, Miranda:2022kzs, Maity:2023rez, refId0, Bose:2022ola, Nguyen:2022zwb}. 

In this paper, we explore a complimentary scenario where DM accumulated through capture in local BDs is annihilated to the sufficiently long-lived mediator. This long-lived mediator escapes the BDs and decays to four neutrinos to produce an observable neutrino flux. To~study the feasibility of probing this DM-induced neutrino flux, we consider the differential neutrino flux obtained or expected from IceCube or IceCube Gen~2.

 In the present work, we study a sample of nearby (within a 20 pc distance) cold and old BDs and look for neutrino excess emission from the direction of these objects using the sky survey data from IceCube and IceCube Gen 2. We do not find any excess from our selected BDs. Two of our candidates show the potential to be detected by IceCube and IceCube Gen 2 but currently lie in the vicinity of 90\% sensitivity limits (Figure~\ref{fig2}).

 As a conservative approach,  we derive the differential flux limits (Figure~\ref{fig2}) as a consequence of the capture of DM particles within the BD and subsequent annihilation into long-lived mediators. The~long-lived mediator, while interacting only feebly within the BD, escapes the celestial object and then decays into neutrinos, leading to a characteristic box-like spectrum. Additionally, considering long-lived mediators, we also study the differential neutrino flux limits for direct spectra to neutrinos (Figure~\ref{fig3}). Our study points out that box-like spectra have more potential to be observed by IceCube and IceCube Gen~2 under certain~conditions.

 Even though box-like spectra would be promising compared to the direct annihilation spectra, our study shows that, at the current epoch, the~expected flux from individual BDs (e.g., Source 9) is a few orders of magnitude weaker than the sensitivity of IceCube or IceCube Gen 2. This work demonstrates a significant improvement in the local BD population. With~our derived number density of local BDs, there are expected to be around 30 BDs within 20 pc, and~by incorporating cumulative emissions from them, we provide a competitive bound with the deepest sensitivity of IceCube Gen 2  (Figure~\ref{fig6}). For~the final impression, we compare the limits on $\sigma_{\chi n}$ obtained from our work with the constraints from direct detection on the DM-nucleon SD cross-section~\cite{CDEX:2018lau, PICO:2019vsc, XENON:2019zpr}, as~well as from DM annihilation from the Sun~\cite{IceCube:2021xzo,refId0}. 

In this study, we focus exclusively on the local population of BDs within 20 pc. As~a future extension, this work further motivates us in the direction of the Galactic Center (GC) population of BDs and examines the relationship between the DM capture rate in the GC BDs, estimated to be around 300 billion~\cite{Leane:2020wob}, and~the diffuse neutrino emission from the Galactic Plane. The~substantial number of BDs in the GC will also play a promising role in seeking robust constraints from current and future neutrino~telescopes.

 In conclusion, this work contributes crucial insights into DM detection strategies for local BDs, suggesting promising avenues for next-generation neutrino telescopes such as IceCube Gen 2 to unveil the mysteries of DM. In~the near future, with~deeper and more sensitive optical and infrared sky surveys such as, e.g.,~via JWST, the~detection of ultracool BDs would be increased, and that would positively improve the sensitivity of our~analysis.

\vspace{6pt} 

\authorcontributions{Conceptualization, formal analysis and draft writing- Done by Pooja Bhattacharjee; Review, suggestions and necessary editing- Done by Francesca Calore.}

\funding{The work of P.B. has been supported by the EOSC Future project, which is co-funded by the European Union Horizon Programme call INFRAEOSC-03-2020, Grant Agreement 101017536.}

\dataavailability{The data would be available upon requesting to the authors.}

\acknowledgments{
P.B.~would like to thank Tarak Maity and Nicholas L.~Rodd for their valuable~suggestions.} 

\conflictsofinterest{The authors declare no conflicts of interest.}
\begin{adjustwidth}{-\extralength}{0cm}
\reftitle{References}

\PublishersNote{}
\end{adjustwidth}
\end{document}